\documentstyle[11pt,newpasp,twoside,epsf]{article}
\def\edcomment#1{\iffalse\marginpar{\raggedright\sl#1\/}\else\relax\fi}
\reversemarginpar
\begin{document}
\title{Clustering of Dark Matter Halos on the Light-cone}
 \author{Yasushi Suto} 
\affil{Department of Physics and Research Center
 for the Early Universe, School of Science, University of
 Tokyo, Tokyo 113-0033, Japan.}

\begin{abstract}
A phenomenological model for the clustering of dark matter halos on the
light-cone is presented.  In particular, an empirical prescription for
the scale-, mass-, and time-dependence of halo biasing is described in
detail.  A comparison of the model predictions against the light-cone
output from the Hubble Volume $N$-body simulation indicates that the
present model is fairly accurate for scale above $\sim 5h^{-1}$Mpc. Then
I argue that the practical limitation in applying this model comes from
the fact that we have not yet fully understood what are clusters of
galaxies, especially at high redshifts. This point of view may turn out
to be too pessimistic after all, but should be kept in mind in
attempting {\it precision cosmology} with clusters of galaxies.
\end{abstract}

\section{Introduction}

The standard picture of cosmological structure formation suggests that
any visible object forms in a gravitational potential of {\it dark
matter halos}.  Therefore, a detailed description of dark halo
clustering is the most basic step toward understanding the clustering of
visible objects in the universe. For this purpose, many theoretical
models for halo clustering have been developed and then tested against
extensive numerical simulations.

First, I will describe our most recent theoretical model for clustering
of dark matter halos (Hamana, Yoshida, Suto \& Evrard 2001b).  In
particular, we focus on their high-redshift clustering where the past
light-cone effect is important. Then I will show that our model
predictions are in good agreement with the result from a light-cone
output of the Hubble volume simulation (Evrard et al. 2001).  Finally I
will discuss a fundamental difficulty in relating the halo model to
clusters of galaxies. My conclusion is that we already have a reliable
empirical model for the halo clustering, but that we need to understand
what are the clusters of galaxies, especially at high redshifts, before
attempting {\it precision cosmology} with clusters of galaxies.

\section{Theoretical model for  two-point correlation functions of 
dark matter halo on the light-cone}

As emphasized by Suto et al.(1999), for instance, observations of
high-redshift objects are carried out only through the past light-cone
defined at $z=0$, and the corresponding theoretical modeling should
properly take account of a variety of physical effects which are briefly
summarized below.

\subsection{Nonlinear gravitational evolution of dark matter density
fluctuations}

 Assuming the cold dark matter (CDM) paradigm, the linear power spectrum
of the mass density fluctuations is computed by solving the Boltzmann
equation for systems of CDM, baryons, photons and (usually massless for
simplicity) neutrinos. The resulting spectrum $P^{\rm R}_{\rm
linear}(k,z)$ in real space is specified by a set of cosmological
parameters including the density parameter $\Omega_0$, the baryon
density parameter $\Omega_b$, and the Hubble constant $h$ in units of
100/km/s/Mpc, and the cosmological constant $\lambda_0$.  Then one can
obtain its nonlinear counterpart in real space, $P^{\rm R}_{\rm
nl}(k,z)$ by adopting a fitting formula of Peacock \& Doods (1996).

\subsection{Empirical model of halo biasing}

 The most important ingredient in describing the clustering of halos is
their biasing properties. The mass-dependent halo bias model was
developed by Mo \& White (1996) on the basis of the extended
Press-Schechter theory.  Subsequently Jing (1998) and Sheth \& Tormen
(1999) improved their bias model so as to more accurately reproduce the
mass-dependence of bias computed from $N$-body simulation results.  We
construct an improved halo bias model of the two-point statistics which
reproduces the scale-dependence of the Taruya \& Suto (2000) bias
correcting the mass-dependent but scale-independent bias of Sheth \&
Tormen (1999) on linear scales as follows:
%%%%%%%%%%%%%%%%%%%%%%%%%%%%%%%%%%%%%%%%%%%%%%%%%%%%%%%%%%%%%%%%%%%%%
\begin{eqnarray}
\label{eq:bhalo_MRz}
b_{\rm halo}(M,R,z) &=&
  b_{\scriptscriptstyle\rm ST}(M,z) 
\left[ 1.0+ b_{\scriptscriptstyle\rm ST}(M,z)\sigma_M(R,z) \right]^{0.15}, \\
b_{\scriptscriptstyle\rm ST}(M,z) &=& 1 + \frac{\nu-1}{\delta_c(z)}
+\frac{0.6}{\delta_c(z)(1+0.9\nu^{0.3})},
\end{eqnarray}
%%%%%%%%%%%%%%%%%%%%%%%%%%%%%%%%%%%%%%%%%%%%%%%%%%%%%%%%%%%%%%%%%%%%%
for $R>2R_{\rm vir}(M,z)$, where $R_{\rm vir}(M,z)$ is the virial radius
of the halo of mass $M$ at $z$. In order to incorporate the halo
exclusion effect approximately, we set $b_{\rm halo}(M,R,z)= 0$ for
$R<2R_{\rm vir}(M,z)$.  In the above expressions, $\sigma_M (R,z)$ is
the mass variance smoothed over the top-hat radius $R\equiv
(3M/4\pi\rho_0)^{1/3}$, $\rho_0$ is the mean density, $\delta_c(z) =
3(12\pi)^{2/3}/20D(z)$, $D(z)$ is the linear growth rate of mass
fluctuations, and $\nu=[\delta_c(z)/\sigma_M (R,z=0)]^2$.

\subsection{Redshift-space distortion}

In linear theory of gravitational evolution of fluctuations, any density
fluctuations induce the corresponding peculiar velocity field, which
results in the systematic distortion of the pattern of distribution of
objects in redshift space (Kaiser 1987). In addition, virialized
nonlinear objects have an isotropic and large velocity dispersion.  This
{\it finger-of-God} effect significantly suppresses the observed
amplitude of correlation on small scales.  With those effects, the
nonlinear power spectrum {\it in redshift space} is given as
%%%%%%%%%%%%%%%%%%%%%%%%%%%%%%%%%%%%%%%%%%%%%%%%%%%%%%%%%%%%%%%%%%%%%
\begin{equation}
P^{\rm S}(k,\mu,z)=P^{\rm R}_{\scriptscriptstyle\rm nl}(k,z)
[1+\beta_{\rm halo}\mu^{2}]^{2}
D_{\rm vel}[k\mu\sigma_{\rm halo}],
\label{nonlinear}
\end{equation}
%%%%%%%%%%%%%%%%%%%%%%%%%%%%%%%%%%%%%%%%%%%%%%%%%%%%%%%%%%%%%%%%%%%%%
where $D_{\rm vel}$ is the Fourier transform of the pairwise peculiar
velocity distribution function (e.g., Magira et al. 2000), $\mu$ is the
direction cosine in $k$-space, $\sigma_{\rm halo}$ is the
one-dimensional {\it pair-wise} velocity dispersion of halos, and
%%%%%%%%%%%%%%%%%%%%%%%%%%%%%%%%%%%%%%%%%%%%%%%%%%%%%%%%%%%%%%%%%%%%%
\begin{equation}
\beta_{\rm halo} \equiv - \frac{1}{b_{\rm halo}} 
\frac{d \ln D(z)}{d \ln z} .
\end{equation}
%%%%%%%%%%%%%%%%%%%%%%%%%%%%%%%%%%%%%%%%%%%%%%%%%%%%%%%%%%%%%%%%%%%%%
While both $\sigma_{\rm halo}$ and $b_{\rm halo}$ depend on the halo
mass $M$, separation $R$, and $z$ in reality, we neglect their
scale-dependence in computing the redshift distortion, and adopt the
halo number-weighted averages:
%%%%%%%%%%%%%%%%%%%%%%%%%%%%%%%%%%%%%%%%%%%%%%%%%%%%%%%%%%%%%%%%%%%%%
\begin{eqnarray}
\label{eq:sigma_halo}
\sigma_{\rm halo}^2(>M,z) &\equiv& 
 \frac{\displaystyle \int_M^\infty 2D^2(z)
\sigma_{\scriptscriptstyle\rm v}^2(M,z=0)\, 
   n_{\scriptscriptstyle\rm J}(M,z) dM}
 {\displaystyle\int_M^\infty n_{\scriptscriptstyle\rm J}(M,z) dM} , \\
\label{eq:b_halo}
b_{\rm halo}(>M,z) &\equiv& 
 \frac{\displaystyle\int_M^\infty b_{\scriptscriptstyle\rm ST}(M,z)\, 
  n_{\scriptscriptstyle\rm J}(M,z) dM}
 {\displaystyle\int_M^\infty n_{\scriptscriptstyle\rm J}(M,z) dM} , 
\end{eqnarray}
%%%%%%%%%%%%%%%%%%%%%%%%%%%%%%%%%%%%%%%%%%%%%%%%%%%%%%%%%%%%%%%%%%%%%
where we adopt the halo mass function $n_{\scriptscriptstyle\rm J}(M,z)$
fitted by Jenkins et al. (2001).  The value of
$\sigma_{\scriptscriptstyle\rm v}(M,z=0)$, the halo center-of-mass
velocity dispersion at $z=0$, is modeled following Yoshida, Sheth \&
Diaferio (2001).  Then our empirical halo bias model can be applied to
the two-point correlation function of halos at $z$ in redshift space as
%%%%%%%%%%%%%%%%%%%%%%%%%%%%%%%%%%%%%%%%%%%%%%%%%%%%%%%%%%%%%%%%%%%%%
\begin{equation}
\label{eq:xihalo_z}
\xi_{\rm halo}(M,R,z) = b^2_{\rm halo}(M,R,z)
\int_0^\infty P^{\rm S}(k,z) \frac{\sin kR}{kR} 
 \frac{k^2 dk}{2\pi^2} .
\end{equation}
%%%%%%%%%%%%%%%%%%%%%%%%%%%%%%%%%%%%%%%%%%%%%%%%%%%%%%%%%%%%%%%%%%%%%

\subsection{Cosmological light-cone effect}

All cosmological observations are carried out on a light-cone, the null
hypersurface of an observer at $z=0$, and not on any constant-time
hypersurface.  Thus clustering amplitude and shape of objects should
naturally evolve even {\it within} the survey volume of a given
observational catalogue. Unless restricting the objects at a narrow bin
of $z$ at the expense of the statistical significance, the proper
understanding of the data requires a theoretical model to take account
of the average over the light cone (Matsubara, Suto, \& Szapudi 1997;
Mataresse et al.  1997; Moscardini et al.  1998; Nakamura, Matsubara, \&
Suto 1998; Yamamoto \& Suto 1999; Suto et al. 1999).  According to the
present prescription, the two-point correlation function of halos on the
light-cone is computed by averaging over the appropriate halo number
density and the comoving volume element between the survey range $z_{\rm
min}<z<z_{\rm max}$:
%%%%%%%%%%%%%%%%%%%%%%%%%%%%%%%%%%%%%%%%%%%%%%%%%%%%%%%%%%%%%%%%%%%
\begin{eqnarray}
\label{eq:xihalolc}
    \xi_{\rm halo}^{\rm\scriptscriptstyle {LC}}(>M,R) 
&=& {
   {\displaystyle \int_M^\infty dM 
     \int_{z_{\rm min}}^{z_{\rm max}} dz 
     {dV_{\rm c} \over dz} ~n_{\scriptscriptstyle\rm J}^2(M,z)
    \xi_{\rm halo}(M,R,z)
    }
\over
    {\displaystyle \int_M^\infty dM 
     \int_{z_{\rm min}}^{z_{\rm max}} dz 
     {dV_{\rm c} \over dz}  ~n_{\scriptscriptstyle\rm J}^2(M,z)
     }
} ,
\end{eqnarray}
%%%%%%%%%%%%%%%%%%%%%%%%%%%%%%%%%%%%%%%%%%%%%%%%%%%%%%%%%%%%%%%%%%%%
where $dV_{\rm c}/dz$ is the comoving volume element per unit solid
angle. While the above expression assumes a mass-limited sample for
simplicity, any observational selection function can be included in the
present model fairly straightforwardly (Hamana, Colombi \& Suto 2001a)
once the the relation between the luminosity of the visible objects and
the mass of the hosting dark matter halos is specified.

\section{Comparison with the light-cone output from the 
 Hubble volume simulation}

%%%%%%%%%%%%%%%%%%%%%%%%%%%%%%%%%%%%%%%%%%%%%%%%%%%%%%%%%%%%%%%%%%%%%%%%%%
\begin{figure}[tbh]
\plotfiddle{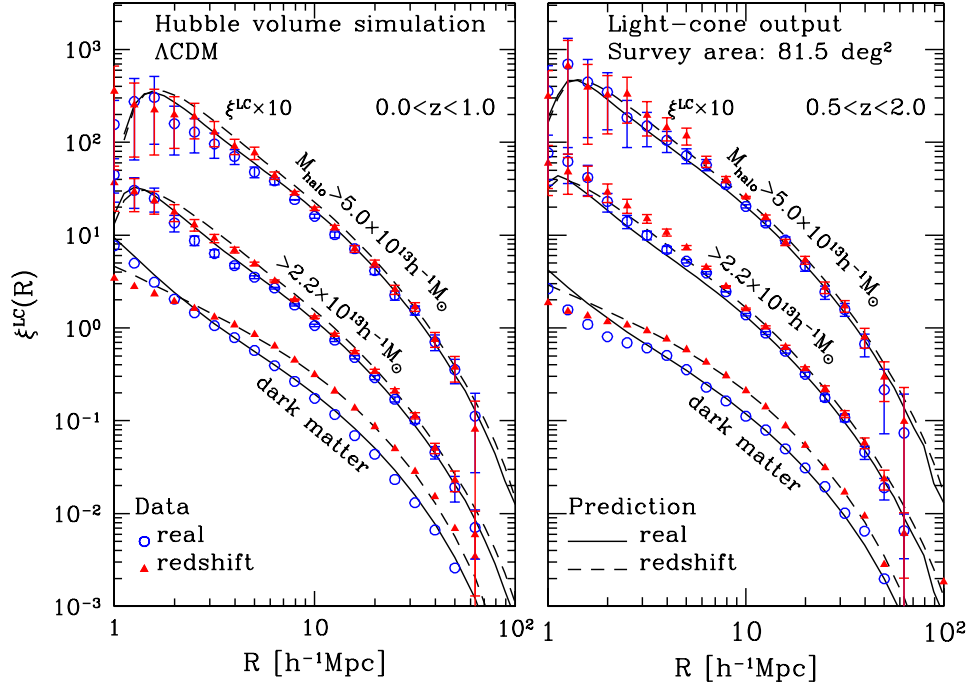}{8.5cm}{0.0}{65.0}{65.0}
{-200.0}{-120.0}
%{-200.0}{-140.0}
\caption{Two-point correlation functions of halos on the light-cone;
simulation results (symbols; open circles and filled triangles for real
and redshift spaces, respectively) and our predictions (solid and dotted
lines for real and redshift spaces, respectively).  The error bars
denote the standard deviation computed from 200 random re-samplings of
the bootstrap method.  The amplitudes of $\xi^{LC}$ for $M_{\rm halo}\ge
5.0\times 10^{13}h^{-1} M_\odot$ are increased by an order of magnitude
for clarity.
\label{fig:haloxilc}}
\end{figure}
%%%%%%%%%%%%%%%%%%%%%%%%%%%%%%%%%%%%%%%%%%%%%%%%%%%%%%%%%%%%%%%%%%%%%%%%%%

Figure 1 compares our model predictions with the clustering of simulated
halos from ``light-cone output'' of the Hubble Volume $\Lambda$CDM
simulation (Evrard et al. 2001) with $\Omega_{\rm b}=0.04$, $\Omega_{\rm
CDM}=0.26$, $\sigma_{8}=0.9$, $\Omega_{\Lambda}=0.7$ and $h=0.7$.  For
the dark matter correlation functions, our model reproduces the
simulation data almost perfectly at $R>3h^{-1}$Mpc (see also Hamana et
al. 2001a). This scale corresponds to the mean particle separation of
this particular simulation, and thus the current simulation
systematically underestimates the real clustering below this scale
especially at $z>0.5$.  Our model and simulation data also show quite
good agreement for dark halos at scales larger than $5h^{-1}$Mpc. Below
that scale, they start to deviate slightly in a complicated fashion
depending on the mass of halo and the redshift range.  This discrepancy
may be ascribed to both the numerical limitations of the current
simulations and our rather simplified model for the halo biasing.
Nevertheless the clustering of {\it clusters} on scales below
$5h^{-1}$Mpc is difficult to determine observationally anyway, and our
model predictions differ from the simulation data only by $\sim 20$
percent at most. Therefore we conclude that in practice our empirical
model provides a successful description of halo clustering on the
light-cone.

\section{Dark matter halos vs. galaxy clusters}

With the successful empirical model of halo clustering, the next natural
question is how to apply it in describing  {\it real}
galaxy clusters. In fact, in my opinion the main obstacle for that
purpose is the lack of the universal definition of clusters. Let me give
some examples that I can easily think of (see Fig.2).

%%%%%%%%%%%%%%%%%%%%%%%%%%%%%%%%%%%%%%%%%%%%%%%%%%%%%%%%%%%%%%%%%%%%%%%%%%
\begin{figure}[tbh]
\plotfiddle{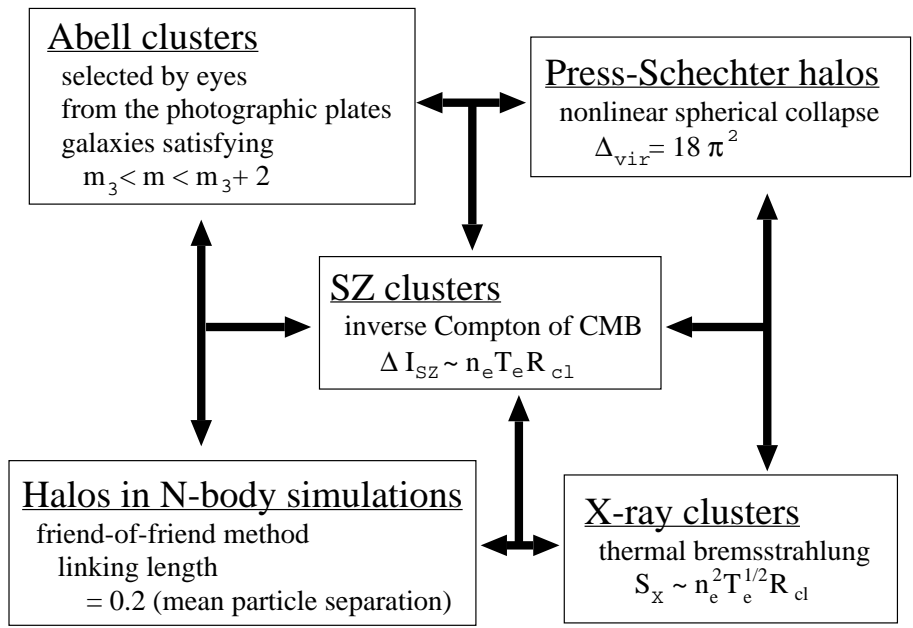}{8.5cm}{0.0}{80.0}{80.0}{-175.0}{-410.0}
\caption{Dark halos -- galaxy clusters connection. There are a variety
 of practical definitions of dark matter halos and clusters of
 galaxies. They are deinitely related, but the one-to-one correspondence
 is unlikely and nothing but a working hypothesis.
\label{fig:halocluster}}
\end{figure}
%%%%%%%%%%%%%%%%%%%%%%%%%%%%%%%%%%%%%%%%%%%%%%%%%%%%%%%%%%%%%%%%%%%%%%%%%%

%%%%%%%%%%%%%%%%%%%%%%%%%%%%%%%%%%%%%%%%%%%%%%%%%%%%%%%%%%%%%%%%%%%%%%
\begin{description}
\item[(i) Press-Schechter halos:] almost all theoretical studies adopt
the definition of dark matter halos according to the nonlinear spherical
model. This is characterized by the mean overdensity of $18\pi^2$ (in
the case of the Einstein - de Sitter universe. The corresponding
expressions in other cosmological models can be also
derived.). Combining this definition with the Press-Schechter theory,
the mass function of the dark halos can be computed analytically. This
makes it fairly straightforward to compare the predictions in this model
with observations, and therefore this definition has been widely studied
in cosmology.
\item[(ii) Halos identified from N-body simulations:] in reality the
gravitationally bound objects in the universe quite often show
significant departure from the spherical symmetry.  Such non-spherical
effects can be directly explored with N-body simulations. Even in this
methodology, the identification of dark halos from the simulated
particle distribution is somewhat arbitrary.  A most conventional method
is the friend-of-friend algorithm. In this algorithm, the linking length
is the only adjustable parameter to controle the resulting halo
sample. Its value is usually set to be 0.2 times the mean particle
separation in the whole simulation which {\it qualitatively} corresponds
to the overdensity of $18\pi^2$ as described above.
\item[(iii) Abell clusters:] until recently most cosmological studies on
galaxy clusters have been based on the Abell catalogue. While this is a
really amazing set of cluster samples, the eye-selection criteria
applied on the Palomar plates are far from objective and cannot be
compared with the above definitions in a quantitative sense.
\item[(iv) X-ray clusters:] the X-ray selection of clusters
significantly improves the reliability of the resulting catalogue due to
the increased signal-to-noise and moreover removes the projection
contamination compared with the optical selection. Nevertheless the
quantitative comparison with halos defined according to (i) or (ii)
requires the knowledge of gas density profile especially in the central
part which fairly dominates the total X-ray emission.
\item[(v) SZ clusters:] the SZ cluster survey is especially important in
probing the high-z universe. In this case, however, the signal is more
sensitive to the temperature profile in clusters than the X-ray
selection, and thus one needs additional information/models for
temperature in order to compare with the X-ray/simulation results.
\end{description}
%%%%%%%%%%%%%%%%%%%%%%%%%%%%%%%%%%%%%%%%%%%%%%%%%%%%%%%%%%%%%%%%%%%%%%
The above consideration raises the importance to examine the systematic
comparison among the resulting {\it cluster/halo} samples selected
differently. In reality, this is a difficult and time-consuming task,
and one might argue that we do not have to worry about such {\it
details} at this point. Such an optimistic point of view may turn out to
be reasonably right after all.  Nevertheless it is still important, at
present, to keep in mind that this simplistic assumption of ``dark halos
= galaxy clusters'' may produce a systematic effect in the detailed
comparison between observational data and theoretical models.

\section{Conclusions}

I have presented a phenomenological model for clustering of dark matter
halos on the light-cone by combining and improving several existing
theoretical models (Hamana et al. 2001b).  One of the most
straightforward and important applications of the current model is to
predict and compare the clustering of X-ray selected clusters.  In doing
so, however, the one-to-one correspondence between dark halos and
observed clusters should be critically examined at some point.  This
assumption is a reasonable working hypothesis, but we need more
quantitative justification or modification to move on to {\it precision
cosmology with clusters}.
 
I am afraid that this problem has not been considered seriously simply
because the agreement between model predictions and available
observations seems already {\it satisfactory}. In fact, since current
viable cosmological models are specified by a set of many {\it
adjustable} parameters, the agreement does not necessarily justify the
underlying assumption. Thus it is dangerous to stop doubting the
unjustified assumption because of the (apparent) success. I hope to
examine these issues in future.

\acknowledgements

I would like to thank Fred Lo for inviting me to this exciting and
enjoyable meeting and also for great hospitality at Taiwan.  The present
work is based on my collaboration with T.Hamana, N.Yoshida, and
A.E.Evrard.  This research was supported in part by the Grant-in-Aid by
the Ministry of Education, Science, Sports and Culture of Japan
(07CE2002, 12640231).

%%%%%%%%%%%%%%%%%%%%%%%%%%%%%%%%%%%%%%%%%%%%%%%%%%%%%%%%%%%%%%%%%%%%%

%%%%%%%%%%%%%%%%%%%%%%%%%%%%%%%%%%%%%%%%%%%%%%%%%%%%%%%%%%%%%%%%%%%%%

\end{document}